%% file: KK_cernpreprint_Feb_2013.tex
\documentclass[ALICE,manyauthors]{cernphprep}
\usepackage{color}

\begin{document}

%
\begin{titlepage}
\PHnumber{2012-337}                 
\PHdate{21 December 2012}              

\title{Charged kaon femtoscopic correlations in pp collisions at $\sqrt{s}=7$~TeV}  

\ShortTitle{Charged kaon femtoscopic correlations in pp collisions at $\sqrt{s}=7$~TeV}  

%
\Collaboration{ALICE Collaboration%
         \thanks{See Appendix~\ref{app:collab} for the list of collaboration
                      members}}
\ShortAuthor{ALICE Collaboration}      

\begin{abstract}
Correlations of two charged identical kaons (${\rm K}^{\rm ch}{\rm K}^{\rm ch}$) are measured in
pp collisions at $\sqrt{s}=7$~TeV by the ALICE experiment
at the Large Hadron Collider (LHC). One-dimensional ${\rm K}^{\rm ch}{\rm K}^{\rm ch}$ correlation 
functions are constructed in three multiplicity and four transverse momentum ranges. 
The ${\rm K}^{\rm ch}{\rm K}^{\rm ch}$ femtoscopic source parameters $R$ and $\lambda$ are extracted.
The ${\rm K}^{\rm ch}{\rm K}^{\rm ch}$ correlations show a slight increase of femtoscopic radii with
increasing multiplicity and a slight decrease of radii with increasing transverse
momentum. These trends are similar to the ones observed for $\pi \pi$ and ${\rm K}^0_{\rm s}{\rm K}^0_{\rm s}$ 
correlations in pp and heavy-ion collisions. However at high multiplicities, there is an indication that the one-dimensional correlation radii for charged kaons are larger than those for pions in contrast to what was observed in heavy-ion collisions at RHIC.
\end{abstract}
\end{titlepage}
\setcounter{page}{2}

\maketitle
\section{\label{sec:sec1}Introduction.}

Extremely high energy densities achieved in heavy-ion collisions at the Large Hadron 
Collider (LHC) may entail the formation of the Quark-Gluon Plasma (QGP), a state
characterized by partonic degrees of freedom  \cite{QGP}.
Studying the QGP is the main goal of the ALICE experiment
(A Large Ion Collider Experiment) \cite{ALICE1}. The system
created in ultrarelativistic pp collisions at LHC energies 
might be similar to the system created in non-central heavy-ion 
collisions because of the large energy deposited 
in the overlapping region and therefore may also manifest a
collective behavior. 
The highly compressed strongly-interacting system is expected to 
undergo longitudinal and transverse expansions.
Experimentally, the expansion and the spatial
extent at decoupling are observable via Bose-Einstein correlations. 

Bose-Einstein correlations of two identical pions at low relative 
momenta were first shown to be sensitive to the spatial scale of 
the emitting source by G. Goldhaber, S. Goldhaber, W. Lee
and A. Pais 50 years ago \cite{GGLP}. The correlation method since developed 
and known at present as ``correlation femtoscopy'' was successfully applied 
to the measurement of the space-time characteristics of particle production 
processes in high energy collisions, especially in heavy-ion collisions 
(see, e.g. \cite{pod89,led04,lis05}). Bose-Einstein correlations of 
identical particles were widely studied in heavy-ion collisions at the Relativistic 
Heavy-Ion Collider (RHIC) \cite{RHIC}, and were found to confirm the hydrodynamic 
type of collective expansion of the fireball created in such collisions.  
In heavy-ion collisions the decrease of the correlation radii with increasing 
particle momentum was usually considered  as a manifestation of a collective 
behavior of the matter created in such collisions \cite{lis05}. 
Event multiplicities reached in 7~TeV pp collisions at the LHC are 
comparable with those measured in peripheral A+A collisions at RHIC, 
making the study of the particle momentum dependence of the correlation 
radii an important test of the collectivity in pp collisions.

The ALICE Collaboration has already studied two-pion correlation radii in pp collisions
at 900~GeV \cite{Aamodt:2010jj} and 7~TeV \cite{Aamodt:2011kd}, and 
${\rm K}^0_{\rm s}{\rm K}^0_{\rm s}$ correlation radii in pp collisions at 7~TeV \cite{Humanic:2011ef}.
Two-pion Bose-Einstein correlations in pp collisions at $\sqrt{s} = 
900$\,GeV and 7~TeV have been successfully described within the EPOS+hydro model \cite{EPOS}. 
It was shown that the hydrodynamic expansion substantially modifies 
the source evolution compared to the ``classical''  EPOS scenario with independent 
decay of flux-tube strings, allowing one to describe the transverse momentum dependence 
of the correlation radii at high multiplicities.

The main motivations for carrying out the present ${\rm K}^{\rm ch}{\rm K}^{\rm ch} ({\rm K}^{\rm +}{\rm K}^{\rm +} + {\rm K}^{\rm -}{\rm K}^{\rm -})$
femtoscopy analysis are: 
1) study the transverse mass, $m_{\rm T}$, dependence of the correlation radii 
(``$m_{\rm T}$-scaling'' is expected to be an additional confirmation of the hydrodynamic 
type of expansion \cite{lis05}), 2) get a clearer signal (kaons are less affected 
by the decay of resonances than pions).

Previous ${\rm K}^{\rm ch}{\rm K}^{\rm ch}$ studies carried out in Pb--Pb collisions at SPS by the NA44 and NA49 
Collaborations \cite{KK_SPS} and in Au--Au collisions at RHIC by the PHENIX Collaboration
 \cite{Afanasiev:2009ii} revealed scaling in transverse mass: the
source sizes versus $m_{\rm T}$ for different particle types ($\pi$, K) fall on the same curve. 

 ${\rm K}^{\rm ch}{\rm K}^{\rm ch}$ studies were performed in a combined data from  
$\alpha\alpha$, pp, and p$\bar{p}$ collisions at ISR by AFS Collaboration 
\cite{KK_ISR}, in e$^+$e$^-$ collisions at LEP by the OPAL and DELPHI Collaborations 
\cite{KK_LEP}, in ep collisions by the ZEUS Collaboration \cite{KK_HERA}. 
Due to statistics limitations, only one-dimensional radii were extracted 
in these experiments, no multiplicity and transverse momentum studies were performed.

In this article we present the measurements of  Bose-Einstein correlations for charged kaons in
pp collisions at $\sqrt{s}=7$~TeV performed by the ALICE Collaboration at the LHC. 
The present study is the first femtoscopic ${\rm K}^{\rm ch}{\rm K}^{\rm ch}$ study to be carried out in pp 
collisions and in more than one multiplicity and pair transverse momentum, $k_{\rm T}$, range.

The paper is organized as follows: in Section~II 
 we describe the ALICE experimental setup and data taking conditions for
the data sample used in this work. In Section~III
we present the correlation measurements 
and the correlation functions. 
In Section~IV
 we show the main results obtained in this work: the one-dimensional radii extracted from the data.
We discuss various observed features and compare the results
with ${\rm K}^0_{\rm s}{\rm K}^0_{\rm s}$ and $\pi\pi$ radii previously
measured by the ALICE Collaboration.
Finally in Section~V
we summarize our results.

\section{Data analysis.}
\label{sec:sec2}
Approximately 300 million minimum-bias events at
 $\sqrt{s} = 7~{\rm TeV}$, recorded in 2010, were analyzed.
The ALICE Time Projection Chamber (TPC) and Inner Tracking
System (ITS) were used for charged particle track reconstruction and the determination of
the primary vertex of the collision.

The TPC identifies charged particles according to their 
ionization trajectories in the $\rm{Ne-CO_2}$ gas. 
The ionization electrons drift up to 2.5 m from the central electrode to the end caps 
to be measured on 159 padrows, grouped into 18 sectors.
The position at which the track crosses the padrow is determined with 
a resolution of 2 mm and 3 mm in the drift and transverse directions, respectively.
The ITS consists of six silicon layers, two innermost Silicon Pixel Detector (SPD) layers, 
two Silicon Drift Detector (SDD) layers, and two outer Silicon Strip Detector (SSD) 
layers, which provide up to six space points for each track. 

 The forward scintillator detectors VZERO were included in the minimum-bias trigger 
and their timing signal was used to reject the beam-gas and beam-halo collisions. 
The minimum-bias trigger required a hit in one of the VZERO counters or in one of 
the two inner layers of the SPD. The VZERO detectors are placed along the beam line at +3~m and
-0.9~m from the nominal interaction point. They cover a region $2.8 < \eta < 5.1$ 
and $-3.7 < \eta < -1.7$.

ALICE provides excellent particle identification capabilities,
using the measurement of specific particle energy loss (${\rm d} E/{\rm d}x$)
in the TPC and the ITS and the time-of-flight ($t_{\rm TOF}$)

information obtained in the Time-Of-Flight (TOF) detector.
The TOF detector is based on Multi-gap Resistive 
Plate Chambers (MRPCs) in a cylindrical configuration at
radius 370-399 cm from the beam axis, with about 153.000 
readout channels of dimension $3.5 \times 2.5$~cm$^2$.  
The start time of the collision (event time zero) is measured by 
the T0 detector, an array of Cherenkov counters located at +350 cm 
and -70 cm along the beam-line. 
If the T0 signal is absent the start time is estimated from the particle 
arrival times at the TOF. 
The overall time-of-flight resolution depends on the TOF timing signal resolution
(better then  100~ps), the accuracy of the reconstructed flight path and the 
uncertainty in the event start time.
The resulting time-of-flight resolution is about 160 ps.

The following criteria were applied for the  event selection:
\begin{itemize}
\item $z$-position of the reconstructed vertex within 
$\pm$10~cm around the geometrical center of the ALICE detector; 
\item at least one particle in the event reconstructed and identified as a charged kaon  
({\it In fact, the correlation signal is constructed from events having at least 
two same-charged kaons (a pair). The one-kaon events do contribute to the mixed background. 
It was verified that including one-kaon event to the mixed background does not change the 
shape of the correlation function.})
\end{itemize}
The criteria for track selection are listed below: 
\begin{itemize}
\item the kaons were selected in the kinematic ranges: $|\eta| < 1.0$ 
and $0.15<p_{\rm T}<1.2$~GeV/$c$; 
\item tracks must include at least 70 space points (or 
clusters) out of a maximum possible number 159 in the TPC, and two 
space points in the ITS (of maximum 6).
\item the quality of a track was determined by the $\chi^2/N$  
 value for the Kalman fit to the constructed position of the TPC clusters 
($N$ is the number of clusters associated with the track); 
the track was rejected if the value
was larger than 4.0 (2 degrees of freedom per cluster);
\item in order to reduce the number of secondary particles 
it was required that the particle trajectory distance from the primary vertex  
was less than 0.2~cm in the transverse plane and less than 0.25~cm in the beam direction.
\end{itemize}
 
Usually the femtoscopic correlation functions 
of identical particles are very sensitive to two-track reconstruction 
effects because particles have close momenta and thus close trajectories.
The ``splitting'' of the tracks means that one track
was reconstructed as two, and ``merging'' means that two different
tracks were reconstructed as one.
   For the correlation structures measured in pp collisions, with characteristic widths 
$\sim$ 0.2 GeV/$c$, track splitting and track merging
in the event reconstruction are small effects,
but we applied the standard femtoscopic double track cuts (see for details
\cite{Aamodt:2011kd}):
\begin{itemize}
\item ``anti-splitting cut'': pairs which share more than 5$\%$ of clusters in the TPC 
   were removed;
\item ``anti-merging cut'': pairs  that are separated by less than 3 cm at the entrance 
  of the TPC were removed.
\end{itemize}
Pair cuts were applied in exactly the same way for real (signal) and mixed  (background) pairs.

In the present analysis the limit $p_{\rm T}< 1.2$~GeV/$c$
for kaon selection on the TPC and TOF signals was used in order to 
ensure a high purity of the kaon sample. Kaons were selected by
requiring that the deviation of the specific (d$E$/d$x$) energy loss in the TPC 
from that calculated with a parametrized Bethe-Bloch formula  
be within some number of sigma standard deviations ($N_{\sigma_{\rm TPC}}$). 
A similar $N_{\sigma_{\rm TOF}}$ method was applied for the particle 
identification in the TOF
using the difference between the measured time-of-flight and the calculated one as
a function of the track length and the particle momentum at each tracking step
and for each particle mass.

More details on the particle identification are given in Ref.~\cite{Aamodt:2011zj},
where it is shown in particular that the fraction efficiency of 
the particles reconstructed by the TPC with associated signal in the 
TOF (the TOF matching efficiency) deviates from 30$\%$ up to 55$\%$ in the $p_{\rm T}$-region 
-- under study in our paper. In the present analysis strict cuts on TPC and TOF signals 
for kaon selection were used in order to provide the better purity of the kaon sample. 
The relative contribution of the kaons from the sample with TOF signal to  the full sample of the identified kaons used in this analysis is about 60$\%$.

In the present analysis strict cuts on TPC and TOF signals for kaon selection
were used in order to provide the better purity of the kaon sample.
If the TOF signal was not available, the following cuts were taken:
\begin{itemize}
\item $N_{\sigma_{\rm TPC}}<1$ for $p<0.35$~GeV/$c$;
\item $N_{\sigma_{\rm TPC}}<2$ for $0.35<p<0.6$~GeV/$c$; 
\item at larger momenta the tracks were rejected because of significant pion 
contamination.
\end{itemize}
If the TOF signal was available, we required that $N_{\sigma_{\rm TOF}}<3$ and
$N_{\sigma_{\rm TPC}}<3$.

\begin{figure}[thb]
\includegraphics[width=8.6cm]{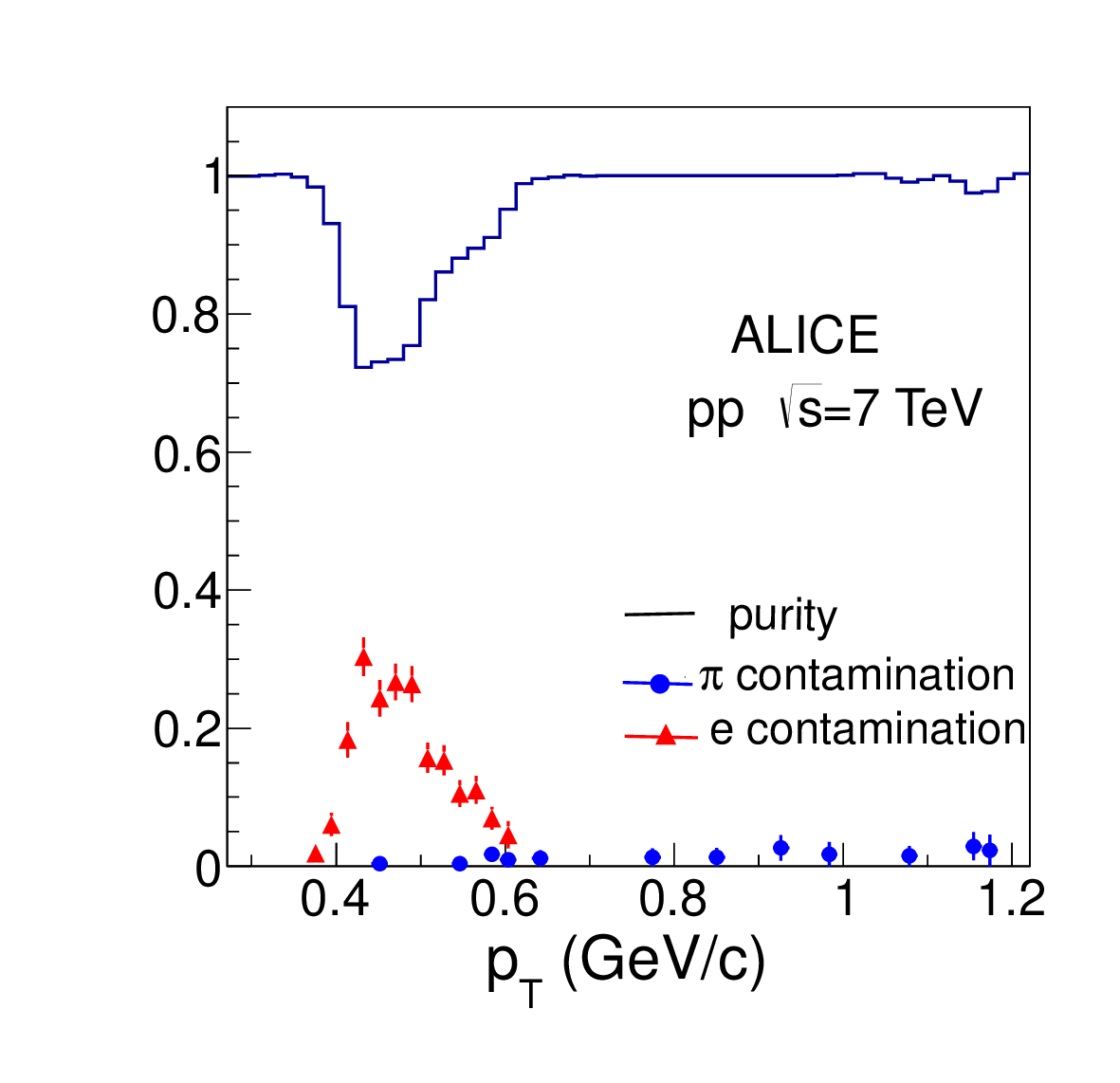}
\caption{Purity and contaminations of selected kaons in pp
collisions at $\sqrt{s}=7$~TeV. }
\label{fig:Purity}
\end{figure}

Figure~\ref{fig:Purity} shows the transverse momentum
dependence of the kaon  purity (the ratio of correctly identified kaons to all
identified ones) obtained with Monte Carlo (PYTHIA) simulations.
The contamination comes mainly from e$^{+}$/e$^{-}$ with maximum 
$\sim 25\%$ for $0.35<p_{\rm T}<0.6$~GeV/$c$ 
and also from pions at the level $\sim 1-3 \%$  for $0.35<p_{\rm T}<1.2$~GeV/$c$. 
In consequence, the probability of selecting
an ee or a $\pi\pi$ pair instead of a KK pair, even in the case 
where the contamination is maximal,
is still rather low, smaller than $\sim 6 \%$. 
The probability of selecting an eK(Ke) pair 
instead of a KK pair can reach $\sim 37 \%$ for $0.35<p_{\rm T}<0.6$~GeV/$c$.
The probability of selecting a $\pi$K(K$\pi$) pair 
instead of a KK pair is less then $\sim 6 \%$ 
for $0.35<p_{\rm T}<1.2$~GeV/$c$. Note that such contamination
modifies only the strength of correlation and not the shape of the correlation function.
Using the Monte Carlo information and also the purity of kaons estimated from the measured TPC d$E$/d$x$
distributions and a realistic single-particle kinematics we have
estimated the purity of kaon pairs at low relative momentum as:
$~p = 0.84 \pm 0.05~\rm{(stat.)} \pm 0.15~\rm{(syst.)},  0.61 \pm 0.03~\rm{(stat.)} \pm 0.12~\rm{(syst.)}, 0.79 \pm 0.04~\rm{(stat.)} \pm 0.07~\rm{(syst.)}, 1.0 \pm 0.05~\rm{(stat.)} \pm 0.05~\rm{(syst.)}$ for pair transverse momentum $k_{\rm T} = |{\bf p}_{{\rm T},1} + {\bf p}_{{\rm T},2}| /2$:
(0.2-0.35), (0.35-0.5), (0.5-0.7), (0.7-1.0)~GeV/$c$, respectively.

\section{Charged kaon correlation functions.}
\label{sec:sec3}
Momentum correlations are usually studied by means of
correlation functions of two or more particles. Specifically, the 
two-particle correlation function 
$CF({\bf p}_1,{\bf p}_2)=A({\bf p}_1,{\bf p}_2)/B({\bf p}_1,{\bf p}_2)$ 
is defined 
as the ratio of the two-particle distribution in a given event
 $A({\bf p}_1,{\bf p}_2)$ to the reference one, $B({\bf p}_1,{\bf p}_2)$, 
 where ${\bf p}_1$ and ${\bf p}_2$ are the momentum vectors of the two particles.
In the present analysis 
the reference distribution is constructed by mixing  
particles of a given class, as described below.

The analysis was performed in three multiplicity ranges 
based on the measured charged-particle multiplicity, $N_{\rm ch}$: 
$(1-11)$, $(12-22)$, $(>22)$, and
in 4 ranges of pair transverse momentum $k_{\rm T}$: 
(0.2-0.35), (0.35-0.5), (0.5-0.7), (0.7-1.0)~GeV/$c$.
Event multiplicity was determined as the number of charged 
particles emitted into the pseudorapidty
range $|\eta|<1$ and transverse momentum range $0.12 < p_{\rm T} < 10$~GeV/$c$. 
For each class of events we calculated the charged-particle pseudorapidity
density ${\rm d}N_{\rm ch}/{\rm d}\eta$ corrected for the detection 
efficiency obtained with Monte Carlo.
The considered event multiplicity ranges $(1-11)$, $(12-22)$, $>22$, 
correspond to mean charged particle densities, 
${\rm d}N_{\rm ch}/{\rm d}\eta$, of 3.2,  8.1 and 17.2, 
respectively, with systematic uncertainties of $\sim 5 \%$. 
 
The numerators and denominators of positive and negative kaon distributions
were summed up before constructing the ratio
(${\rm K}^{+}{\rm K}^{+}$ and ${\rm K}^{-}{\rm K}^{-}$ correlation functions 
were found to coincide within errors thus justifying the procedure). 
The function is normalized to unity in the range $0.5<q_{\rm{inv}}<1.0$~GeV/$c$, where 
$q_{\rm{inv}}= \sqrt{|{\bf q}|^{2} - q_{0}^{2}} $, ${\bf q}={\bf p_1}-{\bf p_2}$, 
and $q_0=E_1-E_2$.
The range for normalization was chosen outside the Bose-Einstein peak.

The correlation function is fitted by a single-Gaussian \cite{BW-Sin}:
\begin{equation}
CF(q_{\rm inv})=\left(1 -\lambda +K(q_{\rm inv})\left( 
\lambda  \exp{\left(-R_{\rm inv}^{2} q^{2}_{\rm inv}\right)}\right)\right)\,D(q_{\rm inv}),
\label{eq:CF}
\end{equation}
where the factor $K(q_{\rm inv})$  is the Coulomb function
integrated over a spherical source of 1~fm.
The function $D(q_{\rm inv})$, ``baseline'', takes into account all non-femtoscopic 
correlations, including the long-range correlations due to 
energy-momentum conservation. The parameters $R_{\rm inv}$ 
and $\lambda$ describe the size of the kaon source, and the correlation 
strength, respectively. 
The $\lambda$-parameter depends also on purity and decreases if the purity is not 100$\%$.
The $R_{\rm inv}$ is measured in the pair rest 
frame.

The baseline was fitted by a standard quadratic polynomial
\begin{equation}
D(q_{\rm inv}) = 1 + a q_{\rm inv}+bq_{\rm inv}^2.
\label{eq:D_P2}
\end{equation}
To estimate the systematic errors due to the fitting procedure, other functions 
with derivatives equal to zero at $q_{\rm inv} = 0$ 
were also employed, such as:
\begin{equation}
D(q_{\rm inv}) = \sqrt{1 + a q_{\rm inv}^{2}+bq_{\rm inv}^4}
\label{eq:D_sqrt}
\end{equation}
and the Gaussian:
\begin{equation}
D(q_{\rm inv}) = b (1+exp(-a q_{\rm inv}^{2})).
\label{eq:D_Gaus}
\end{equation}

The PERUGIA-2011 tune \cite{PERUGIA2011} of the Monte Carlo event generator 
PYTHIA \cite{PYTHIA} describes well the kaon spectra in pp collisions
at LHC energies. Therefore, it was used to simulate 
the correlation function without the Bose-Einstein effect.

\section{Results and Discussion.}
\label{sec:sec4}
\begin{figure}[thb]
\includegraphics[width=8.6cm]{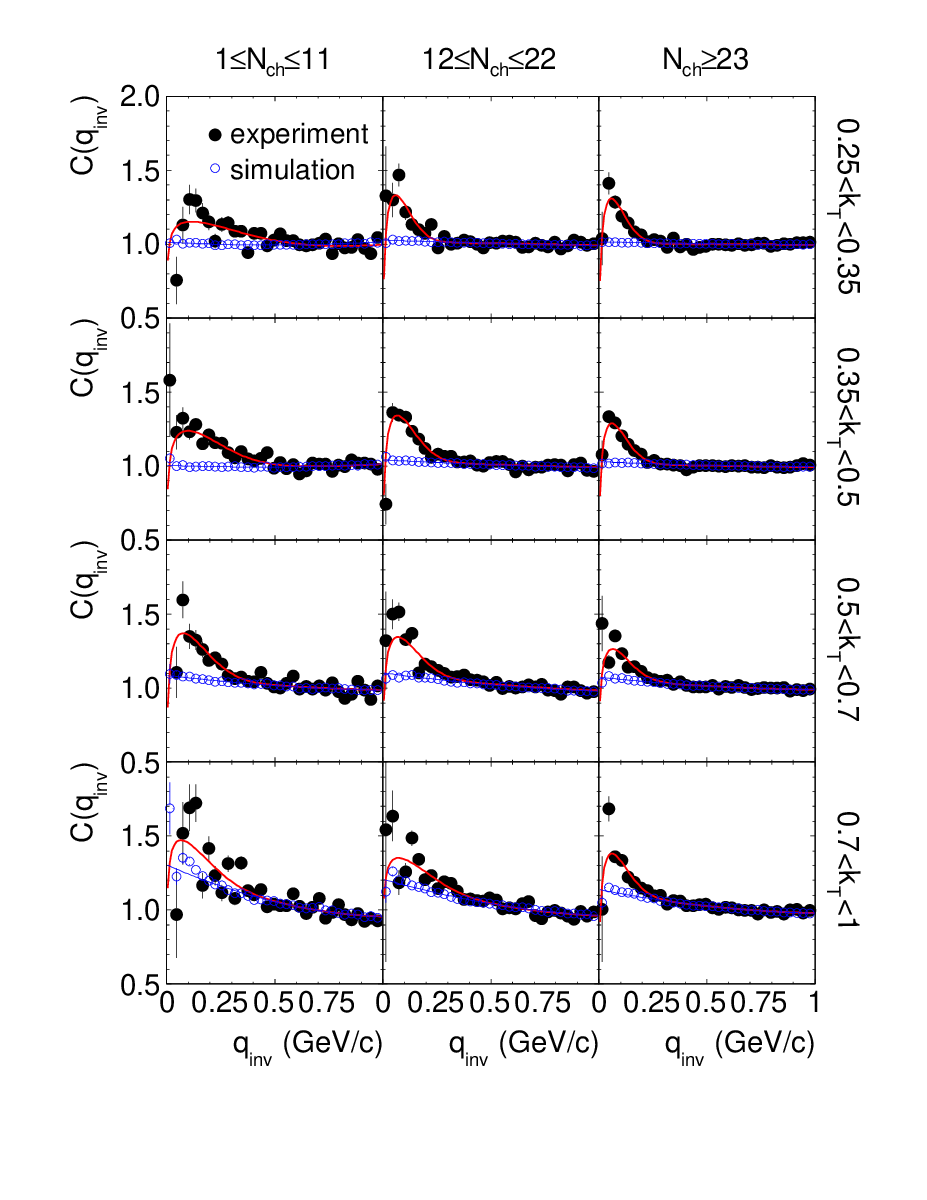}
\caption{
Correlation functions versus $q_{inv}$ for identical kaons from pp
collisions at $\sqrt{s}=7$~TeV (solid circles)
and those obtained with PERUGIA-2011 (open circles). Positive and negative
kaon  pairs are combined. The three columns represent 
the samples with different charged-particle multiplicities:
 $(1-11)$, $(12-22)$, $(>22)$, the four rows represent
the four pair transverse momentum ranges:
(0.2-0.35), (0.35-0.5), (0.5-0.7), (0.7-1.0)~GeV/$c$.
The lines going
through the points represent the Gaussian fits discussed in the text.
}
\label{fig:CFs_KK}
\end{figure}

\begin{figure}[thb]
\includegraphics[width=8.6cm]{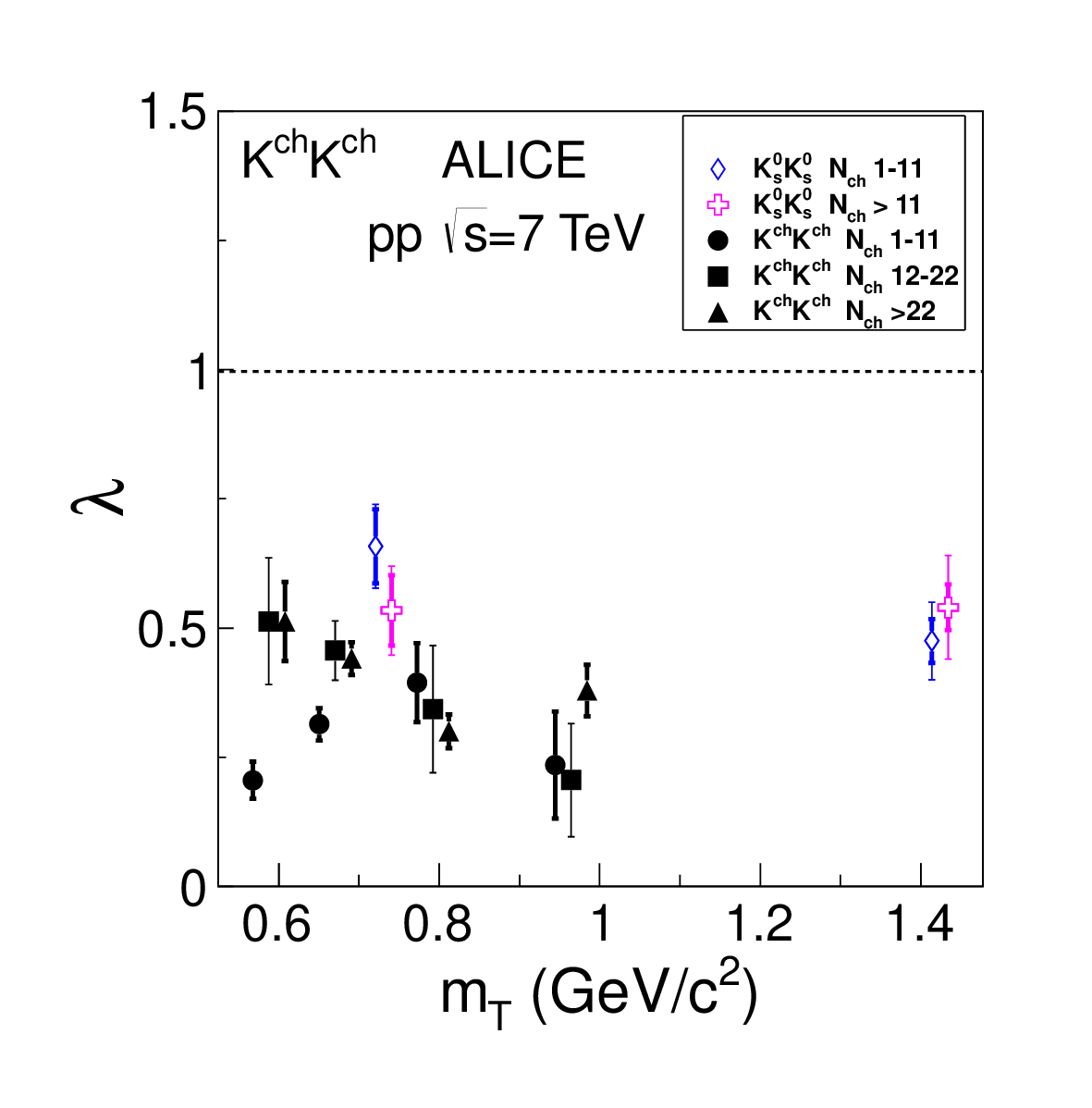}
\caption{
$\lambda$-parameters of ${\rm K}^{\rm ch}{\rm K}^{\rm ch}$ versus $m_{\rm T}$ 
extracted by fitting correlation functions shown in 
Fig.~\ref{fig:CFs_KK} to Eq.~(\ref{eq:CF}) and the baseline to Eq.~(\ref{eq:D_P2}). 
For comparison the 
${\rm K}^0_{\rm s}{\rm K}^0_{\rm s}$ \cite{Humanic:2011ef} 
$\lambda$-parameters measured by ALICE in 7~TeV
pp collisions are also shown.
Statistical (darker lines) and total errors are shown.
The points corresponding to the second and third multiplicity bins
are offset by 0.03~GeV/$c^2$ for clarity.
}
\label{fig:Lam_KK_K0K0}
\end{figure}

\begin{figure}[thb]
\includegraphics[width=8.6cm]{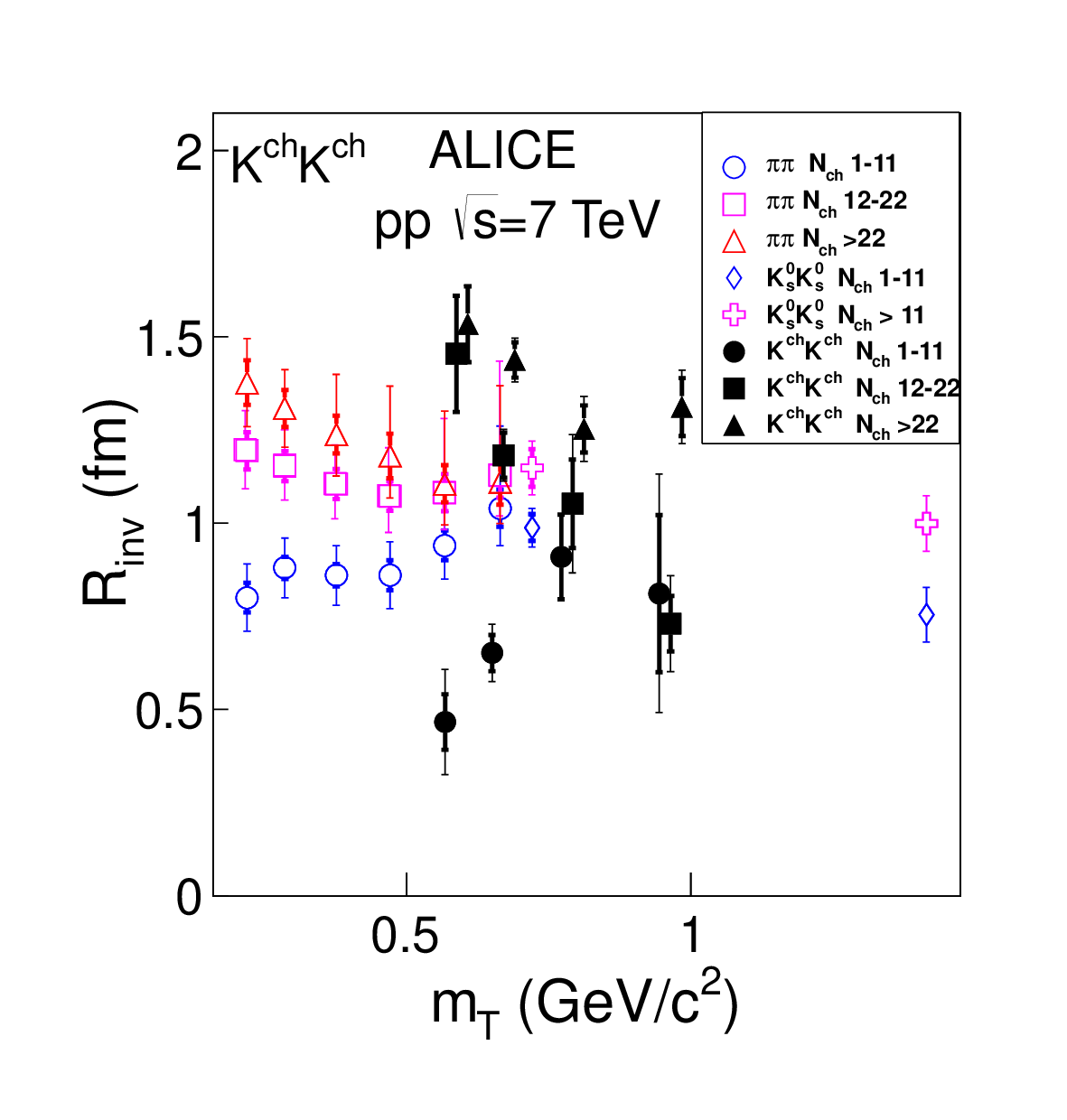}
\caption{
One-dimensional charged kaon radii versus $m_{\rm T}$ 
extracted by fitting correlation functions shown in 
Fig.~\ref{fig:CFs_KK} to Eq.~(\ref{eq:CF}) and the baseline to Eq.~(\ref{eq:D_P2}). 
For comparison the $\pi\pi$\cite{Aamodt:2011kd} 
and  ${\rm K}^0_{\rm s}{\rm K}^0_{\rm s}$ \cite{Humanic:2011ef} 
radii measured by ALICE in 7~TeV
pp collisions are also shown.
Statistical (darker lines) and total errors are shown.
The points corresponding to the second and third multiplicity bins
are offset by 0.03~GeV/$c$$^2$ for clarity.
}
\label{fig:R_KK_K0K0}
\end{figure}

Figure~\ref{fig:CFs_KK} presents the experimental two-kaon correlation functions
 and those obtained from a simulation using PERUGIA-2011 (open circles)
as a function of the invariant pair relative momentum.
As one can see, the Monte Carlo simulation reproduces well the experimental correlation function at 
large $q_{\rm inv}$, i.e. the long-range correlations. 
The model does not contain the Bose-Einstein effect, 
so the enhancement at low $q_{\rm inv}$ is due to non-femtoscopic correlations in PYTHIA,
probably arising from mini-jets.
The baseline points, obtained from PERUGIA-2011 were fitted to Eq.~(\ref{eq:D_P2}). 
The parameters $a$ and $b$ were used in the fitting of the experimental points by Eq.~(\ref{eq:CF}).
The same method was used to model the baseline for the ALICE 
$\pi \pi$ correlation studies in 0.9 \cite{Aamodt:2010jj} and 7~TeV \cite{Aamodt:2011kd}
pp collisions. 
Figures~\ref{fig:Lam_KK_K0K0}-\ref{fig:R_KK_K0K0} and Table~\ref{tab:R_Lam} present 
the one-dimensional $\lambda$-parameters and
Gaussian radii versus $m_{\rm T} = \sqrt{k_{\rm T}^2+m_{\rm K}^{2}}$ 
including statistical and systematical errors.

In order to estimate the systematic error from the choice of baseline functional form 
we repeated the fitting  procedure using the baseline fitted with Eqs.~(\ref{eq:D_sqrt}-\ref{eq:D_Gaus}).
The radii obtained in the three ways differ by less than 4 $\%$ 
at low multiplicities (or $k_{\rm T}$)
and by up to 10$\%$ at high  multiplicities (or $k_{\rm T}$).
The systematic errors estimated from varying the $q_{\rm inv}$ fit range are 
below 2~$\%$ and up to 15$\%$ at low and high multiplicities (or $k_{\rm T}$) bins respectively.
The systematic errors from splitting and merging effects were estimated by using 
different start points for the fit of correlation function: 0.03 and 0.06~GeV/$c$, and its
are about 2-6$\%$.
The systematic error connected to the Coulomb function in Eq.~(\ref{eq:CF})
was calculated in the following way: at first, the radius of the spherical source 
was taken equal to $1$~fm, then the fitting procedure was repeated using 
these radii $\pm 3 \sigma_R$ (where $\sigma_R$ is the total error)
as the argument of the Coulomb function. The obtained systematic error is about 2-4$\%$.

The $m_{\rm T}$ dependence of $\lambda$ shown in Fig.~\ref{fig:Lam_KK_K0K0}
demonstrates that $\lambda$ varies within the range $\sim 0.3-0.5$
(except the first point at lowest multiplicity 
and lowest $k_{\rm T}$, which is $\sim 0.2$). 
As seen in Fig.~\ref{fig:Lam_KK_K0K0}, the $\lambda$-parameters for 
${\rm K}^{\rm ch}{\rm K}^{\rm ch}$ 
are generally smaller than those for ${\rm K}^{0}_{s}{\rm K}^{0}_{s}$ \cite{Humanic:2011ef}. 
There are several reason for the $\lambda$-parameter to be less than the ideal 
case of unity.
Possible causes may be such as a partially coherent source, a contribution to
the observed kaons from decays of long-lived resonances, a deviation from
the Gaussian parameterization due to a mixture of sources with different radii 
(see e.g. \cite{lambda}), or a particle misidentification. The latter cause influences mostly the 
${\rm K}^{\rm ch}{\rm K}^{\rm ch}$ sample in the momentum range $0.4<p_{\rm T}<0.6$~GeV/$c$ 
(Fig.~\ref{fig:Purity}).
For the future, it seems desirable to improve the PID and thus increase the purity and
to improve the description of non-femtoscopic correlations using different models.

The ${\rm K}^{\rm ch}{\rm K}^{\rm ch}$ correlation radii in Fig.~\ref{fig:R_KK_K0K0} 
show an increase with multiplicity 
in agreement with the $\pi \pi$ radii at 900~GeV \cite{Aamodt:2010jj}, and 7~TeV 
\cite{Aamodt:2011kd},  and the ${\rm K}^0_{\rm s}{\rm K}^0_{\rm s}$ 
radii \cite{Humanic:2011ef}, as it was observed for $\pi \pi$ correlations 
in heavy-ion collisions \cite{Aamodt:2011mr}.
These radii also decrease with increasing $m_{\rm T}$ 
for the large multiplicity bins
$N_{\rm ch}$ $(12-22)$ and $N_{\rm ch}$ $(>22)$. Such a tendency was found 
for pions \cite{Aamodt:2011kd} and neutral kaons 
${\rm K}^0_{\rm s}{\rm K}^0_{\rm s}$ \cite{Humanic:2011ef} in pp collisions,
and pions in heavy-ion collisions at LHC energies \cite{Aamodt:2011mr}.
In the low multiplicity bin $N_{\rm ch}$ $(1-11)$
charged kaons show a completely different $k_{\rm T}$-dependence of the radii:
these radii increase with $k_{\rm T}$. This effect is qualitatively
similar to that of pions \cite{Aamodt:2011kd}.

It was observed that $\lambda$-parameters (Fig.~\ref{fig:Lam_KK_K0K0})
are correlated with the radii (Fig.~\ref{fig:R_KK_K0K0}).
Such a correlation can result from the non-perfect fit results 
if the fit quality depends on $m_{\rm T}$.
The reasons of such a dependence are:  
non ideal description of the baseline by PERUGIA-2011, 
especially at large $k_{\rm T}$, 
non-Gaussian shape  of the source  due to resonance contribution 
and non-spherical shape of the source.
The last two points mean that our one-dimensional Gaussian
fit is only an approximate description of the source.   
In pp  collisions the effect of this non-Gaussian shape of the correlation function 
due to different sizes in the 'x-y-z' directions plays a more important role than in  
heavy-ion collisions. This requires a detailed 3D analysis, which 
is foreseen for ${\rm K}^{\rm ch}{\rm K}^{\rm ch}$ correlation functions with 
the new large set of data recorded by the ALICE Collaboration in 2011 and 2012.

The $m_{\rm T}$ dependence of the radii in heavy-ion collisions was 
interpreted as the manifestation of the 
strong collective hydrodynamic expansion of the created matter
\cite{lis05}. The observed similar behavior in pp collisions, shown in Fig.\ref{fig:R_KK_K0K0}, 
has some specific features:  
1) at low multiplicity the radii increase with $k_{\rm T}$, 
2) there is no distinct $m_{\rm T}$ scaling: the
kaon radii seem to be larger than the pion ones. 
The model calculations performed in \cite{EPOS} can
successfully describe the different behavior of pion correlation radii 
in low and high multiplicity bins, suggesting that 
the contribution of the hydrodynamic phase is negligible in 
low-multiplicity events, while for events with high multiplicity, it is 
substantial.

As shown in \cite{Kisiel:2010xy}, due to the small size of the created system
in pp collisions, the flow of resonances 
may play a significant role in large multiplicity bins, where essential hydrodynamic
collective flow is expected \cite{EPOS}. According to   
simple chemical model calculations \cite{Humanic:2011ef},  
the influence of this flow should be relatively smaller for kaons than for pions, 
leading to the effect that the kaon radii can be larger than the pion ones.                              
The measured ${\rm K}^{\rm ch}{\rm K}^{\rm ch}$ correlation radii 
displayed in Fig.\ref{fig:R_KK_K0K0}
support such an hypothesis, however a detailed theoretical 
study is needed.

\begin{table*}
\caption{
${\rm K}^{\rm ch}{\rm K}^{\rm ch}$ source parameters vs. $k_{T}$ for 
$\sqrt{s}$=7~TeV pp collisions. Statistical and
systematic errors are listed.
}

\begin{tabular}{lcccccc}\hline
 $k_{\rm T}$ range & $  N_{\rm ch}  $ & ${\rm d}N_{\rm ch}/{\rm d}\eta$ & $   <k_{\rm T}>  $ & $  \lambda   $  &  $ R _{\rm inv}  $  \\
(GeV/$c$) &                 &    & (GeV/$c$) &                &   (fm)  \\
\hline
0.20-0.35  &  1-11 & 3.2 &   0.28 $\pm$ 0.04 &  0.20 $\pm$ 0.04 $\pm$ 0.03    &   0.47 $\pm$ 0.07 $\pm$ 0.12 \\
0.35-0.50  &  1-11 & 3.2 & 0.42 $\pm$ 0.05 &  0.31 $\pm$ 0.03 $\pm$ 0.02    &   0.65 $\pm$ 0.05 $\pm$ 0.06 \\
0.50-0.70 &  1-11 &  3.2 & 0.59 $\pm$ 0.06 &  0.39 $\pm$ 0.08 $\pm$ 0.03    &   0.91 $\pm$ 0.10 $\pm$ 0.06 \\
0.07-1.00   &  1-11 &  3.2 & 0.80 $\pm$ 0.08 &  0.23 $\pm$ 0.10 $\pm$ 0.20    &   0.81 $\pm$ 0.21 $\pm$ 0.24 \\
\hline
0.20-0.35  &  12-22 & 8.1 & 0.28 $\pm$ 0.04 &  0.51 $\pm$ 0.12 $\pm$ 0.03    &   1.45 $\pm$ 0.15 $\pm$ 0.02 \\
0.35-0.50  &  12-22 & 8.1 & 0.42 $\pm$ 0.05 &  0.46 $\pm$ 0.04 $\pm$ 0.04    &   1.18 $\pm$ 0.06 $\pm$ 0.03 \\
0.50-0.70 &  12-22 &  8.1 & 0.59 $\pm$ 0.06 &  0.34 $\pm$ 0.07 $\pm$ 0.10    &   1.05 $\pm$ 0.12 $\pm$ 0.14 \\
0.70-1.00   &  12-22 & 8.1 & 0.80 $\pm$ 0.08 &  0.21 $\pm$ 0.04 $\pm$ 0.10    &   0.73 $\pm$ 0.07 $\pm$ 0.10 \\
\hline
0.20-0.35  &  $>$22 &  17.2 & 0.28 $\pm$ 0.04 &  0.51 $\pm$ 0.08 $\pm$ 0.03    &   1.53 $\pm$ 0.10 $\pm$ 0.02 \\
0.35-0.50  &  $>$22 &  17.2 & 0.42 $\pm$ 0.05 &  0.44 $\pm$ 0.03 $\pm$ 0.04    &   1.44 $\pm$ 0.04 $\pm$ 0.03 \\
0.50-0.70 &  $>$22 &   17.2 & 0.59 $\pm$ 0.06 &  0.30 $\pm$ 0.03 $\pm$ 0.04    &   1.25 $\pm$ 0.06 $\pm$ 0.06 \\
0.70-1.00   &  $>$22 & 17.2 & 0.80 $\pm$ 0.08 &  0.37 $\pm$ 0.05 $\pm$ 0.06    &   1.31 $\pm$ 0.08 $\pm$ 0.08 \\
\hline
\end{tabular}
\label{tab:R_Lam}
\end{table*}

\maketitle
\section{Summary.}
\label{sec5}
The ALICE Collaboration has measured charged kaon correlation
functions in pp collisions at $\sqrt{s}$ = 7~TeV at the LHC.
In agreement with the previous measurements in pp and heavy-ion collisions at lower energies, the extracted correlation radii $R_{\rm inv}$ increase with the event multiplicity and decrease with the pair transverse mass/momentum. The novel features are some hints to the increase of the radii with $m_{\rm T}$ in the low-multiplicity bin and to the fact that kaon radii are larger than the pion ones. These peculiarities deserve further experimental and theoretical studies.

\newenvironment{acknowledgement}{\relax}{\relax}
\begin{acknowledgement}
\section{Acknowledgements}
\input{acknowledgements_oct2012.tex}    
\end{acknowledgement}
\newpage
\clearpage
%
%
\appendix
\section{The ALICE Collaboration}
\label{app:collab}
\input{authors-preprint.tex}{\relax}{\relax}

\end{document}

%% file: acknowledgements_oct2012.tex
The ALICE collaboration would like to thank all its engineers and technicians for their invaluable contributions to the construction of the experiment and the CERN accelerator teams for the outstanding performance of the LHC complex.
\\
The ALICE collaboration acknowledges the following funding agencies for their support in building and
running the ALICE detector:
 \\
State Committee of Science, Calouste Gulbenkian Foundation from
Lisbon and Swiss Fonds Kidagan, Armenia;
 \\
Conselho Nacional de Desenvolvimento Cient\'{\i}fico e Tecnol\'{o}gico (CNPq), Financiadora de Estudos e Projetos (FINEP),
Funda\c{c}\~{a}o de Amparo \`{a} Pesquisa do Estado de S\~{a}o Paulo (FAPESP);
 \\
National Natural Science Foundation of China (NSFC), the Chinese Ministry of Education (CMOE)
and the Ministry of Science and Technology of China (MSTC);
 \\
Ministry of Education and Youth of the Czech Republic;
 \\
Danish Natural Science Research Council, the Carlsberg Foundation and the Danish National Research Foundation;
 \\
The European Research Council under the European Community's Seventh Framework Programme;
 \\
Helsinki Institute of Physics and the Academy of Finland;
 \\
French CNRS-IN2P3, the `Region Pays de Loire', `Region Alsace', `Region Auvergne' and CEA, France;
 \\
German BMBF and the Helmholtz Association;
\\
General Secretariat for Research and Technology, Ministry of
Development, Greece;
\\
Hungarian OTKA and National Office for Research and Technology (NKTH);
 \\
Department of Atomic Energy and Department of Science and Technology of the Government of India;
 \\
Istituto Nazionale di Fisica Nucleare (INFN) and Centro Fermi -
Museo Storico della Fisica e Centro Studi e Ricerche "Enrico
Fermi", Italy;
 \\
MEXT Grant-in-Aid for Specially Promoted Research, Ja\-pan;
 \\
Joint Institute for Nuclear Research, Dubna;
 \\
National Research Foundation of Korea (NRF);
 \\
CONACYT, DGAPA, M\'{e}xico, ALFA-EC and the HELEN Program (High-Energy physics Latin-American--European Network);
 \\
Stichting voor Fundamenteel Onderzoek der Materie (FOM) and the Nederlandse Organisatie voor Wetenschappelijk Onderzoek (NWO), Netherlands;
 \\
Research Council of Norway (NFR);
 \\
Polish Ministry of Science and Higher Education;
 \\
National Authority for Scientific Research - NASR (Autoritatea Na\c{t}ional\u{a} pentru Cercetare \c{S}tiin\c{t}ific\u{a} - ANCS);
 \\
Ministry of Education and Science of Russian Federation,
International Science and Technology Center, Russian Academy of
Sciences, Russian Federal Agency of Atomic Energy, Russian Federal
Agency for Science and Innovations and CERN-INTAS;
 \\
Ministry of Education of Slovakia;
 \\
Department of Science and Technology, South Africa;
 \\
CIEMAT, EELA, Ministerio de Educaci\'{o}n y Ciencia of Spain, Xunta de Galicia (Conseller\'{\i}a de Educaci\'{o}n),
CEA\-DEN, Cubaenerg\'{\i}a, Cuba, and IAEA (International Atomic Energy Agency);
 \\
Swedish Research Council (VR) and Knut $\&$ Alice Wallenberg
Foundation (KAW);
 \\
Ukraine Ministry of Education and Science;
 \\
United Kingdom Science and Technology Facilities Council (STFC);
 \\
The United States Department of Energy, the United States National
Science Foundation, the State of Texas, and the State of Ohio.

%% file: authors-preprint.tex
\begingroup
\small
\begin{flushleft}
B.~Abelev\Irefn{org1234}\And
J.~Adam\Irefn{org1274}\And
D.~Adamov\'{a}\Irefn{org1283}\And
A.M.~Adare\Irefn{org1260}\And
M.M.~Aggarwal\Irefn{org1157}\And
G.~Aglieri~Rinella\Irefn{org1192}\And
M.~Agnello\Irefn{org1313}\textsuperscript{,}\Irefn{org1017688}\And
A.G.~Agocs\Irefn{org1143}\And
A.~Agostinelli\Irefn{org1132}\And
Z.~Ahammed\Irefn{org1225}
% ...
\renewcommand\labelenumi{\textsuperscript{\theenumi}~}
\section*{Affiliation notes}
\renewcommand\theenumi{\roman{enumi}}
\begin{Authlist}
\item \Adef{0}Deceased
\item \Adef{M.V.Lomonosov Moscow State University, D.V.Skobeltsyn Institute of Nuclear Physics, Moscow, Russia}Also at: M.V.Lomonosov Moscow State University, D.V.Skobeltsyn Institute of Nuclear Physics, Moscow, Russia
\item \Adef{University of Belgrade, Faculty of Physics and Vinvca Institute of Nuclear Sciences, Belgrade, Serbia}Also at: University of Belgrade, Faculty of Physics and Vinvca Institute of Nuclear Sciences, Belgrade, Serbia
\end{Authlist}
\section*{Collaboration Institutes}
\renewcommand\theenumi{\arabic{enumi}~}
\begin{Authlist}
\item \Idef{org1332}A. I. Alikhanyan National Science Laboratory (Yerevan Physics Institute) Foundation, Yerevan, Armenia
\item \Idef{org1279}Benem\'{e}rita Universidad Aut\'{o}noma de Puebla, Puebla, Mexico
\item \Idef{org1220}Bogolyubov Institute for Theoretical Physics, Kiev, Ukraine
\item \Idef{org20959}Bose Institute, Department of Physics and Centre for Astroparticle Physics and Space Science (CAPSS), Kolkata, India
\item \Idef{org1262}Budker Institute for Nuclear Physics, Novosibirsk, Russia
\item \Idef{org1292}California Polytechnic State University, San Luis Obispo, California, United States
\item \Idef{org1329}Central China Normal University, Wuhan, China
\item \Idef{org14939}Centre de Calcul de l'IN2P3, Villeurbanne, France
\item \Idef{org1197}Centro de Aplicaciones Tecnol\'{o}gicas y Desarrollo Nuclear (CEADEN), Havana, Cuba
\item \Idef{org1242}Centro de Investigaciones Energ\'{e}ticas Medioambientales y Tecnol\'{o}gicas (CIEMAT), Madrid, Spain
\item \Idef{org1244}Centro de Investigaci\'{o}n y de Estudios Avanzados (CINVESTAV), Mexico City and M\'{e}rida, Mexico
\item \Idef{org1335}Centro Fermi - Museo Storico della Fisica e Centro Studi e Ricerche ``Enrico Fermi'', Rome, Italy
\item \Idef{org17347}Chicago State University, Chicago, United States
\item \Idef{org1288}Commissariat \`{a} l'Energie Atomique, IRFU, Saclay, France
\item \Idef{org15782}COMSATS Institute of Information Technology (CIIT), Islamabad, Pakistan
\item \Idef{org1294}Departamento de F\'{\i}sica de Part\'{\i}culas and IGFAE, Universidad de Santiago de Compostela, Santiago de Compostela, Spain
\item \Idef{org1106}Department of Physics Aligarh Muslim University, Aligarh, India
\item \Idef{org1121}Department of Physics and Technology, University of Bergen, Bergen, Norway
\item \Idef{org1162}Department of Physics, Ohio State University, Columbus, Ohio, United States
\item \Idef{org1300}Department of Physics, Sejong University, Seoul, South Korea
\item \Idef{org1268}Department of Physics, University of Oslo, Oslo, Norway
\item \Idef{org1315}Dipartimento di Fisica dell'Universit\`{a} and Sezione INFN, Trieste, Italy
\item \Idef{org1145}Dipartimento di Fisica dell'Universit\`{a} and Sezione INFN, Cagliari, Italy
\item \Idef{org1312}Dipartimento di Fisica dell'Universit\`{a} and Sezione INFN, Turin, Italy
\item \Idef{org1285}Dipartimento di Fisica dell'Universit\`{a} `La Sapienza' and Sezione INFN, Rome, Italy
\item \Idef{org1154}Dipartimento di Fisica e Astronomia dell'Universit\`{a} and Sezione INFN, Catania, Italy
\item \Idef{org1132}Dipartimento di Fisica e Astronomia dell'Universit\`{a} and Sezione INFN, Bologna, Italy
\item \Idef{org1270}Dipartimento di Fisica e Astronomia dell'Universit\`{a} and Sezione INFN, Padova, Italy
\item \Idef{org1290}Dipartimento di Fisica `E.R.~Caianiello' dell'Universit\`{a} and Gruppo Collegato INFN, Salerno, Italy
\item \Idef{org1103}Dipartimento di Scienze e Innovazione Tecnologica dell'Universit\`{a} del Piemonte Orientale and Gruppo Collegato INFN, Alessandria, Italy
\item \Idef{org1114}Dipartimento Interateneo di Fisica `M.~Merlin' and Sezione INFN, Bari, Italy
\item \Idef{org1237}Division of Experimental High Energy Physics, University of Lund, Lund, Sweden
\item \Idef{org1192}European Organization for Nuclear Research (CERN), Geneva, Switzerland
\item \Idef{org1227}Fachhochschule K\"{o}ln, K\"{o}ln, Germany
\item \Idef{org1122}Faculty of Engineering, Bergen University College, Bergen, Norway
\item \Idef{org1136}Faculty of Mathematics, Physics and Informatics, Comenius University, Bratislava, Slovakia
\item \Idef{org1274}Faculty of Nuclear Sciences and Physical Engineering, Czech Technical University in Prague, Prague, Czech Republic
\item \Idef{org1229}Faculty of Science, P.J.~\v{S}af\'{a}rik University, Ko\v{s}ice, Slovakia
\item \Idef{org1184}Frankfurt Institute for Advanced Studies, Johann Wolfgang Goethe-Universit\"{a}t Frankfurt, Frankfurt, Germany
\item \Idef{org1215}Gangneung-Wonju National University, Gangneung, South Korea
\item \Idef{org20958}Gauhati University, Department of Physics, Guwahati, India
\item \Idef{org1212}Helsinki Institute of Physics (HIP) and University of Jyv\"{a}skyl\"{a}, Jyv\"{a}skyl\"{a}, Finland
\item \Idef{org1203}Hiroshima University, Hiroshima, Japan
\item \Idef{org1254}Indian Institute of Technology Bombay (IIT), Mumbai, India
\item \Idef{org36378}Indian Institute of Technology Indore, Indore, India (IITI)
\item \Idef{org1266}Institut de Physique Nucl\'{e}aire d'Orsay (IPNO), Universit\'{e} Paris-Sud, CNRS-IN2P3, Orsay, France
\item \Idef{org1277}Institute for High Energy Physics, Protvino, Russia
\item \Idef{org1249}Institute for Nuclear Research, Academy of Sciences, Moscow, Russia
\item \Idef{org1320}Nikhef, National Institute for Subatomic Physics and Institute for Subatomic Physics of Utrecht University, Utrecht, Netherlands
\item \Idef{org1250}Institute for Theoretical and Experimental Physics, Moscow, Russia
\item \Idef{org1230}Institute of Experimental Physics, Slovak Academy of Sciences, Ko\v{s}ice, Slovakia
\item \Idef{org1127}Institute of Physics, Bhubaneswar, India
\item \Idef{org1275}Institute of Physics, Academy of Sciences of the Czech Republic, Prague, Czech Republic
\item \Idef{org1139}Institute of Space Sciences (ISS), Bucharest, Romania
\item \Idef{org27399}Institut f\"{u}r Informatik, Johann Wolfgang Goethe-Universit\"{a}t Frankfurt, Frankfurt, Germany
\item \Idef{org1185}Institut f\"{u}r Kernphysik, Johann Wolfgang Goethe-Universit\"{a}t Frankfurt, Frankfurt, Germany
\item \Idef{org1177}Institut f\"{u}r Kernphysik, Technische Universit\"{a}t Darmstadt, Darmstadt, Germany
\item \Idef{org1256}Institut f\"{u}r Kernphysik, Westf\"{a}lische Wilhelms-Universit\"{a}t M\"{u}nster, M\"{u}nster, Germany
\item \Idef{org1246}Instituto de Ciencias Nucleares, Universidad Nacional Aut\'{o}noma de M\'{e}xico, Mexico City, Mexico
\item \Idef{org1247}Instituto de F\'{\i}sica, Universidad Nacional Aut\'{o}noma de M\'{e}xico, Mexico City, Mexico
\item \Idef{org23333}Institut of Theoretical Physics, University of Wroclaw
\item \Idef{org1308}Institut Pluridisciplinaire Hubert Curien (IPHC), Universit\'{e} de Strasbourg, CNRS-IN2P3, Strasbourg, France
\item \Idef{org1182}Joint Institute for Nuclear Research (JINR), Dubna, Russia
\item \Idef{org1199}Kirchhoff-Institut f\"{u}r Physik, Ruprecht-Karls-Universit\"{a}t Heidelberg, Heidelberg, Germany
\item \Idef{org20954}Korea Institute of Science and Technology Information, Daejeon, South Korea
\item \Idef{org1017642}KTO Karatay University, Konya, Turkey
\item \Idef{org1160}Laboratoire de Physique Corpusculaire (LPC), Clermont Universit\'{e}, Universit\'{e} Blaise Pascal, CNRS--IN2P3, Clermont-Ferrand, France
\item \Idef{org1194}Laboratoire de Physique Subatomique et de Cosmologie (LPSC), Universit\'{e} Joseph Fourier, CNRS-IN2P3, Institut Polytechnique de Grenoble, Grenoble, France
\item \Idef{org1187}Laboratori Nazionali di Frascati, INFN, Frascati, Italy
\item \Idef{org1232}Laboratori Nazionali di Legnaro, INFN, Legnaro, Italy
\item \Idef{org1125}Lawrence Berkeley National Laboratory, Berkeley, California, United States
\item \Idef{org1234}Lawrence Livermore National Laboratory, Livermore, California, United States
\item \Idef{org1251}Moscow Engineering Physics Institute, Moscow, Russia
\item \Idef{org1322}National Centre for Nuclear Studies, Warsaw, Poland
\item \Idef{org1140}National Institute for Physics and Nuclear Engineering, Bucharest, Romania
\item \Idef{org1017626}National Institute of Science Education and Research, Bhubaneswar, India
\item \Idef{org1165}Niels Bohr Institute, University of Copenhagen, Copenhagen, Denmark
\item \Idef{org1109}Nikhef, National Institute for Subatomic Physics, Amsterdam, Netherlands
\item \Idef{org1283}Nuclear Physics Institute, Academy of Sciences of the Czech Republic, \v{R}e\v{z} u Prahy, Czech Republic
\item \Idef{org1264}Oak Ridge National Laboratory, Oak Ridge, Tennessee, United States
\item \Idef{org1189}Petersburg Nuclear Physics Institute, Gatchina, Russia
\item \Idef{org1170}Physics Department, Creighton University, Omaha, Nebraska, United States
\item \Idef{org1157}Physics Department, Panjab University, Chandigarh, India
\item \Idef{org1112}Physics Department, University of Athens, Athens, Greece
\item \Idef{org1152}Physics Department, University of Cape Town and  iThemba LABS, National Research Foundation, Somerset West, South Africa
\item \Idef{org1209}Physics Department, University of Jammu, Jammu, India
\item \Idef{org1207}Physics Department, University of Rajasthan, Jaipur, India
\item \Idef{org1200}Physikalisches Institut, Ruprecht-Karls-Universit\"{a}t Heidelberg, Heidelberg, Germany
\item \Idef{org1017688}Politecnico di Torino, Turin, Italy
\item \Idef{org1325}Purdue University, West Lafayette, Indiana, United States
\item \Idef{org1281}Pusan National University, Pusan, South Korea
\item \Idef{org1176}Research Division and ExtreMe Matter Institute EMMI, GSI Helmholtzzentrum f\"ur Schwerionenforschung, Darmstadt, Germany
\item \Idef{org1334}Rudjer Bo\v{s}kovi\'{c} Institute, Zagreb, Croatia
\item \Idef{org1298}Russian Federal Nuclear Center (VNIIEF), Sarov, Russia
\item \Idef{org1252}Russian Research Centre Kurchatov Institute, Moscow, Russia
\item \Idef{org1224}Saha Institute of Nuclear Physics, Kolkata, India
\item \Idef{org1130}School of Physics and Astronomy, University of Birmingham, Birmingham, United Kingdom
\item \Idef{org1338}Secci\'{o}n F\'{\i}sica, Departamento de Ciencias, Pontificia Universidad Cat\'{o}lica del Per\'{u}, Lima, Peru
\item \Idef{org1313}Sezione INFN, Turin, Italy
\item \Idef{org1115}Sezione INFN, Bari, Italy
\item \Idef{org1133}Sezione INFN, Bologna, Italy
\item \Idef{org1146}Sezione INFN, Cagliari, Italy
\item \Idef{org1155}Sezione INFN, Catania, Italy
\item \Idef{org1271}Sezione INFN, Padova, Italy
\item \Idef{org1286}Sezione INFN, Rome, Italy
\item \Idef{org1316}Sezione INFN, Trieste, Italy
\item \Idef{org36377}Nuclear Physics Group, STFC Daresbury Laboratory, Daresbury, United Kingdom
\item \Idef{org1258}SUBATECH, Ecole des Mines de Nantes, Universit\'{e} de Nantes, CNRS-IN2P3, Nantes, France
\item \Idef{org35706}Suranaree University of Technology, Nakhon Ratchasima, Thailand
\item \Idef{org1304}Technical University of Split FESB, Split, Croatia
\item \Idef{org1168}The Henryk Niewodniczanski Institute of Nuclear Physics, Polish Academy of Sciences, Cracow, Poland
\item \Idef{org17361}The University of Texas at Austin, Physics Department, Austin, TX, United States
\item \Idef{org1173}Universidad Aut\'{o}noma de Sinaloa, Culiac\'{a}n, Mexico
\item \Idef{org1296}Universidade de S\~{a}o Paulo (USP), S\~{a}o Paulo, Brazil
\item \Idef{org1149}Universidade Estadual de Campinas (UNICAMP), Campinas, Brazil
\item \Idef{org1239}Universit\'{e} de Lyon, Universit\'{e} Lyon 1, CNRS/IN2P3, IPN-Lyon, Villeurbanne, France
\item \Idef{org1205}University of Houston, Houston, Texas, United States
\item \Idef{org20371}University of Technology and Austrian Academy of Sciences, Vienna, Austria
\item \Idef{org1222}University of Tennessee, Knoxville, Tennessee, United States
\item \Idef{org1310}University of Tokyo, Tokyo, Japan
\item \Idef{org1318}University of Tsukuba, Tsukuba, Japan
\item \Idef{org21360}Eberhard Karls Universit\"{a}t T\"{u}bingen, T\"{u}bingen, Germany
\item \Idef{org1225}Variable Energy Cyclotron Centre, Kolkata, India
\item \Idef{org1306}V.~Fock Institute for Physics, St. Petersburg State University, St. Petersburg, Russia
\item \Idef{org1323}Warsaw University of Technology, Warsaw, Poland
\item \Idef{org1179}Wayne State University, Detroit, Michigan, United States
\item \Idef{org1143}Wigner Research Centre for Physics, Hungarian Academy of Sciences, Budapest, Hungary
\item \Idef{org1260}Yale University, New Haven, Connecticut, United States
\item \Idef{org15649}Yildiz Technical University, Istanbul, Turkey
\item \Idef{org1301}Yonsei University, Seoul, South Korea
\item \Idef{org1327}Zentrum f\"{u}r Technologietransfer und Telekommunikation (ZTT), Fachhochschule Worms, Worms, Germany
\end{Authlist}
\endgroup